\documentclass{article}
\usepackage{spconf,amsmath,graphicx}

\usepackage{array}
\usepackage{amsfonts}
\usepackage{tikz}
\usepackage{amsmath}
\usepackage{comment}
\usepackage{color}
\usepackage{booktabs}
\usepackage{xspace}
\usepackage{amssymb}
\usepackage{amsfonts}           
\usepackage{mathrsfs} 
\usepackage{subfigure}
\usepackage{multirow}
\usepackage{booktabs}
\usepackage{graphicx}
\usepackage{enumitem}
\usepackage{mathtools}
\usepackage{cite}
\usepackage{siunitx}
\usepackage{pgfplots}
\pgfplotsset{width=6cm}

\def\presec{\vspace{-1em}}
\def\postsec{\vspace{-0.5em}}

\newcommand{\zengwei}[1]{[{ \color{red} Zengwei: #1}]}

\title{Delay-penalized transducer for low-latency streaming ASR}
\name{Wei Kang\textsuperscript{*,1}, Zengwei Yao\textsuperscript{*,1}, Fangjun Kuang\textsuperscript{1}, Liyong Guo\textsuperscript{1}, \thanks{* stands for equal contribution}}
\myname{Xiaoyu Yang\textsuperscript{1}, Long lin\textsuperscript{1}, Piotr Żelasko\textsuperscript{2}, Daniel Povey\textsuperscript{1}}

\address{\textsuperscript{1} Xiaomi Corp., Beijing, China \:\:\: \textsuperscript{2} Meaning.Team Inc, USA \\
\footnotesize{\texttt{\{kangwei1, yaozengwei, dpovey\}@xiaomi.com, pzelasko@meaning.team}} \vspace{-1em}}

\begin{document}
% \ninept
%
\maketitle
\begin{abstract}

In streaming automatic speech recognition (ASR), it is desirable to reduce latency as much as possible while having minimum impact on recognition accuracy. 
Although a few existing methods are able to achieve this goal, they are difficult to implement due to their dependency on external alignments. In this paper, we propose a simple way to penalize symbol delay in transducer model, 
%a popular modeling choice for ASR, 
so that we can balance the trade-off between symbol delay and accuracy for streaming models without external alignments. 
Specifically, our method adds a small constant times (T/2 - t), where T is the number of frames and t is the current frame, to all the non-blank log-probabilities (after normalization) that are fed into the two dimensional transducer recursion.
For both streaming Conformer models and unidirectional long short-term memory (LSTM) models, experimental results show that it can significantly reduce the symbol delay with an acceptable performance degradation.
Our method achieves similar delay-accuracy trade-off to the previously published FastEmit, 
% In a plot of WER versus delay, our method looks very similar to the previously published FastEmit, 
but we believe our method is preferable because it has a better justification: it is equivalent to penalizing the average symbol delay. Our work is open-sourced and publicly available\footnote{https://github.com/k2-fsa/k2}.
% FastEmit is a modification of the backpropagation that multiplies a particular part of the gradients of transducer by one plus a small constant. 
%Our WER-versus-delay experiments are on streaming Conformer models (i.e. trained with block-triangular masks that limit the future context the model can see) and unidirectional LSTM models.

%The crux of streaming Automatic Speech Recognition (ASR) is to achieve both high recognition precision and low latency.  Whilst transducer has demonstrated its superior performance and essentially streaming advantage, its objective function only concentrates on the total log-probability over all alignments but ignore the consequent symbol delay of different alignments. 
%When applied on streaming models, the unconstrained transducer objective function favours those alignments emitting symbols later to access more future contexts, which leads to a high emission latency. In this paper, we propose the delay-penalized transducer, which is able to balance the performance and emission latency in a simple and efficient way. The only modification is to add time-dependent scores on the log-probabilities of emitting symbols, without the need of any auxiliary token-time alignment information. We demonstrate the effectiveness of the proposed delay-penalized transducer by conducting extensive experiments on two typical streaming encoders: streaming Conformer with limited future contexts and Long Short-Term Memory (LSTM) without any future context. Experimental results shows that ** 

\end{abstract}
\presec
\section{Introduction}
\postsec
End-to-end models have achieved remarkable success in Automatic Speech Recognition (ASR). As a prominent example, transducer~\cite{transducer, transformer-transducer, pruned-rnnt} has gained more and more popularity for real-time ASR system development, because it is
naturally streaming and demonstrates superior performance.
However, one limitation of transducer is that it focuses on maximizing the total log-probability over all alignments but ignores their specific symbol delays. We hypothesize that the streaming model would augment those alignments emitting symbols later to access more contexts for better performance, leading to higher emission latency in practical ASR application. 

There are several classical methods~\cite{cd-ctc-smbr, fast_and_accurate, emit_word_timing, min_latency, align_restrict} to reduce the model latency by constraining the alignments between the frames and transcriptions based on the alignment references generated from external models.  Whilst this type of methods achieve good trade-offs between accuracy and latency, it suffers from two limitations: 1) the model performance heavily depends on the precision of the 
reference alignments; 2) it defeats the advantage of end-to-end model training since it requires an extra frame-level token-time alignments. 

To address these limitations, another line of research ~\cite{fastemit, On-Device, self-alignment} tends to regularize the objective function in a sequence-level manner. A prominent example is FastEmit~\cite{fastemit}, which encourages the model to emit symbols earlier by scaling up the derivatives of emitting non-blank tokens in backpropagation. Another work named Self alignment~\cite{self-alignment} proposes to boost the log-probability of the alignment that is one frame to the left of the Viterbi forced-alignment, which requires an extra recursion with a time complexity of $\mathcal{O}(T \times U)$ to obtain the Viterbi forced-alignment, where $T$ and $U$ are the lengths of frame sequence and token sequence respectively.  Our method, like FastEmit~\cite{fastemit}, is simple to implement, but we are able to provide a more detailed demonstration explaining why our method would cause alignment times to change.

In this paper, we propose a novel method of delay penalization for transducer which is able to balance the trade-off between symbol delay and accuracy for streaming models in a simple and efficient way. Different from FastEmit~\cite{fastemit} that directly changes the derivatives, we modify the log-probabilities of emitting symbols by adding a small
constant $\lambda$ times the frame offsets relative to middle frame. We mathematically prove that it is approximately equivalent to adding a regularization term that aims to decrease the averaged symbol delay on the regular transducer objective function. 
% which regularizes the transducer objective function for low-latency streaming ASR. 
% Specifically, we add a score inversely proportional to the symbol delay on different alignment paths, by modifying the log-probabilities of emitting symbols during the forward-backward process according to the frame indexes.
% We mathematically prove that it is approximately equivalent to adding a regularization term on the regular transducer objective function, which aims to encourage the low-delay alignments and discourage the high-delay alignments. 

The main contributions of this paper are: 
% \begin{itemize}[itemsep=1.8pt,topsep=1.8pt,parsep=0pt]
\begin{itemize}[leftmargin=*,topsep=1.8pt,parsep=0pt,itemsep=1.8pt,]
 \item We propose the delay-penalized transducer, which penalizes the symbol delay without extra token-time alignment. 
 \item We provide a detailed proof why it can encourage the low-delay alignments and penalize the high-delay alignments.
 \item We show that a tunable trade-off between latency and accuracy can be achieved by adjusting the hyperparameter $\lambda$.
\end{itemize}
\label{sec:intro}

\presec
\section{Transducer}
\postsec
\begin{figure}
    \centering
    \includegraphics[width=7cm]{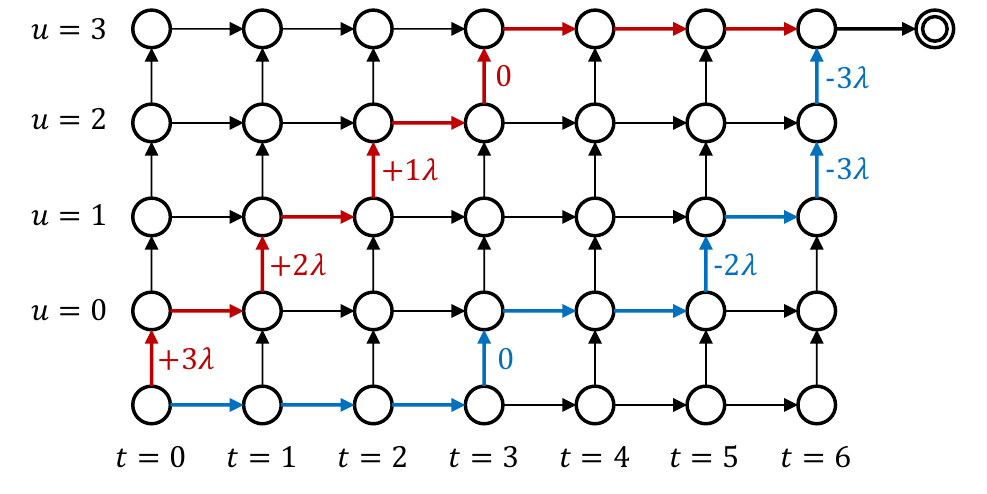}
    \caption{Delay penalized transducer lattice.}
    \label{fig:lattice}
    \vspace{-5mm}
\end{figure}

Let $\mathbf{x} = \{x_t\}_0^{T-1}$ be a sequence of $T$ parameterized input feature frames. Let $\mathbf{y} = \{0 \le y_u < V \}_0^{U-1}$ be a sequence of $U$ transcript tokens, where $V$ is the vocabulary size containing the blank token $\varnothing$.
As shown in Figure~\ref{fig:lattice}, transducer~\cite{transducer} learns alignments between these two sequences $\mathbf{x}$ and $\mathbf{y}$ with different lengths.
The vertical transition leaving node $(t, u)$ represents emitting non-blank token $y_{u+1}$ with the log-probility $y(t, u)$, while the horizontal transition represents emitting blank token $\varnothing$ with the log-probility $\varnothing(t, u)$.

% Specifically, the node at position $(t, u)$ represents emitting tokens $y_{0 \dots u}$ after seeing features $x_{0 \dots t}$. The vertical transition leaving node $(t, u)$ represents emitting non-blank token $y_{u+1}$, while the horizontal transition represents emitting blank token $\varnothing$.

% Typically, transducer consists of three network modules: encoder, decoder and joiner~\cite{transducer}. The encoder learns acoustic features and the decoder functions as a language model. The joiner models the log-probability of the vertical transitions $y(t, u)$ and horizontal transitions $\varnothing(t, u)$ based on the outputs from the encoder and decoder.

The objective function of transducer is to maximize the total log-probability $\mathcal{L}$ over all alignment paths:
\begin{equation}
\vspace{-1mm}
\label{eq:rnnt_L}
\mathcal{L} = \log \sum_i \exp(s_i),
\vspace{-1mm}
\end{equation}
where $s_i$ is the log-probability of path $i$ summing over all contained transitions. The forward-backward algorithm~\cite{transducer} is usually employed to calculate $\mathcal{L}$ in an efficient manner. Let $\alpha(t, u)$ be the log-probability at node $(t, u)$, which represents emitting tokens $y_{0 \dots u}$ after seeing features $x_{0 \dots t}$. Then $\alpha(t, u)$ could be calculated recursively as:
\begin{equation}
\vspace{-1mm}
\label{eq:rnnt_forward}
\begin{aligned}
\alpha(t, u) = \text{LogAdd}(\alpha(t, u-1) + y(t, u-1), \\ \alpha(t-1, u) + \varnothing(t-1, u)),
\end{aligned}
\vspace{-1mm}
\end{equation}
where $\text{LogAdd}$ is defined as:
\begin{equation}
\text{LogAdd}(a, b) = \log{(e^a + e^b)}.
\end{equation}
Herein, $\alpha(0, 0)$ is initialized as 0. The total log-probability of over all alignments path $\mathcal{L}$ is:
\begin{equation}
\label{eq:rnnt_total_p}
\mathcal{L} = \alpha(T-1, U) + \varnothing(T-1, U).
\end{equation}

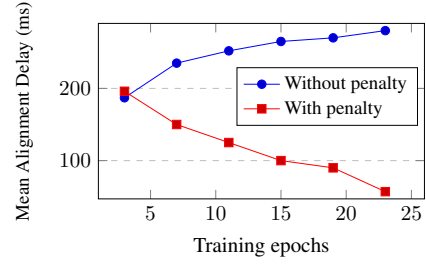
\begin{figure}
\vspace{-1em}
\centering
    \begin{tikzpicture}[scale=0.8]
        \begin{axis}[
            xlabel={Training epochs},
            ylabel={\small{Mean Alignment Delay (ms)}},
            xmin=1, xmax=26,
            ymin=47, ymax=290,
            %ticklabel style = {font=\small},
            % xtick={1,5,10,15,20,25},
            % ytick={-50, 0, 50, 100, 150, 200, 250, 300},
            ymajorgrids=true,
            width=7cm,
            height=4.5cm,
            grid style=dashed,
            legend cell align={left},
            legend style={
              font=\small,
              at={(0.98, 0.750)}
            },
        ]
        % \addplot+ [
        %     every mark/.append style={scale=1}]
        %     coordinates {
        %     (3,-33)(7,-22)(11,-23)(15,-26)(19,-28)(23,-29)
        %     };
        \addplot+ [
            every mark/.append style={scale=1}]
            coordinates {
            (3,187)(7,235)(11,252)(15,265)(19,270)(23,280)
            };
        \addplot+ [
            every mark/.append style={scale=1}]
            coordinates {
            (3,196)(7,150)(11,125)(15,100)(19,90)(23,57)
            };
        \legend{Without penalty,With penalty}
        \end{axis}
    \end{tikzpicture}
\caption{Mean alignment delay of streaming model during training.}
\label{fig:training_delay}
\vspace{-5mm}
\end{figure}

One limitation of transducer is that it is optimized to maximize the total log-probability $\mathcal{L}$ over all alignments, regardless of their respective symbol delays.
As shown in Figure~\ref{fig:lattice}, the blue alignment emitting symbols later has a higher delay compared with the red alignment.
Unlike non-streaming model that could access full contexts in an utterance, the streaming model tends to concentrate on those alignments emitting symbols later, such as the blue alignment in Figure~\ref{fig:lattice}, thus to access more future contexts for a better recognition performance.
The blue line in Figure~\ref{fig:training_delay} presents the
mean alignment delay of the streaming Conformer, which constantly increase as the training goes on.

%which shows that the delay gets larger and larger while the model optimization goes.
%mean alignment delay during training of the streaming Conformer model trained with the transducer loss.

\label{sec:motivation}

\presec
\section{Delay-penalized Transducer}
\postsec

% revised-2022-10-23

%% As we pointed out in Section~\ref{sec:motivation}, the regular transducer tends to maximize the total log-probability $\mathcal{L}$ over all alignments without considering the different symbol emitting delay of each path. The basic idea of our proposed delay-penalized transducer is to encourage the low-delay alignments while optimizing the transducers.

To penalize symbol delay, we (conceptually) add an extra term in the loss function:
\begin{equation}
\label{eq:L_aug}
\mathcal{L}_{\text{aug}} = \mathcal{L} + \mathcal{L}_{\text{delay}}.
\end{equation} 
Herein, $\mathcal{L}_{\text{delay}}$ represents the scaled weighted average delay score over all alignments, which is formulated as:

\begin{equation}
\label{eq:L_delay}
\mathcal{L}_{\text{delay}} = \lambda \sum_i d_i w_i, 
\end{equation}
where $d_i$ is the delay score of alignment $i$, $\lambda$ is a scaling hyper-parameter, and $w_i$ is the path weight:
\begin{equation}
\label{eq:w_delay}
w_i = \frac{\partial \mathcal{L}}{\partial s_i} = \frac{\exp{(s_i)}}{\sum_i \exp{(s_i)}}.
\end{equation}
Herein, the sum of weight $w_i$ over all alignments is 1. 
We can get the derivatives of $\mathcal{L}_{\text{aug}}$ with respect to $s_i$ as:
\begin{equation}
\label{eq:d_aug_s}
\frac{\partial \mathcal{L}_{\text{aug}}}{\partial s_i} 
= \frac{\partial \mathcal{L}}{\partial s_i} + \frac{\partial \mathcal{L}_{\text{delay}}}{\partial s_i} 
\end{equation}
From \eqref{eq:L_delay} and \eqref{eq:w_delay}, we can get:
\begin{equation}
\label{eq:d_delay_s}
\frac{\partial \mathcal{L}_{\text{delay}}}{\partial s_i} = \lambda \left(\frac{d_i\exp{(s_i)}}{\sum_i \exp{(s_i)}} - \frac{d_i(\exp{(s_i)})^2}{(\sum_i \exp{(s_i)})^2}\right),
\end{equation}
which can be rearranged as:
\begin{equation}
\label{eq:d_delay_s_2}
\frac{\partial \mathcal{L}_{\text{delay}}}{\partial s_i} = \lambda \frac{(d_i - d_{\text{avg}})\exp{(s_i)}}{\sum_i \exp{(s_i)}},
\end{equation}
where $d_{\text{avg}}$ is:
\begin{equation}
\label{eq:d_avg}
d_{avg} = \sum_i d_i w_i. 
\end{equation}
From \eqref{eq:w_delay}, \eqref{eq:d_aug_s} and \eqref{eq:d_delay_s_2} we can get:
\begin{equation}
\label{eq:d_aug_s_simple}
\frac{\partial \mathcal{L}_{\text{aug}}}{\partial s_i} 
= \frac{(1 + \lambda (d_i - d_{\text{avg}}))\exp{(s_i)}}{\sum_i \exp{(s_i)}}.
\end{equation}
For a small $\lambda$, $1 + \lambda (d_i - d_{\text{avg}})$ is close to $\exp{(\lambda (d_i - d_{\text{avg}}))}$, we can approximate \eqref{eq:d_aug_s_simple} as:
\begin{equation}
\label{eq:d_aug_s_round}
\frac{\partial L_{\text{aug}}}{\partial s_i} \approx \frac{\exp{(\lambda (d_i - d_{\text{avg}}) + s_i)}}{\sum_i \exp{(s_i)}}.
\end{equation}
According to \eqref{eq:d_delay_s_2} and \eqref{eq:d_avg}, the sum of derivative $ \frac{\partial \mathcal{L}_{\text{delay}}}{\partial s_i}$ over all alignments is 0. By plugging in $\frac{\partial \mathcal{L}}{\partial s_i}$ from \eqref{eq:w_delay}, we can get:
\begin{equation}
\vspace{-1mm}
\label{eq:sum_d_aug_s}
\sum_i \frac{\partial \mathcal{L}_{\text{aug}}}{\partial s_i} = \sum_i \frac{\partial \mathcal{L}}{\partial s_i} + \sum_i \frac{\partial \mathcal{L}_{\text{delay}}}{\partial s_i} = 1.
\vspace{-1mm}
\end{equation}
Then we can equivalently normalize \eqref{eq:d_aug_s_round} as:
\begin{equation}
\label{eq:d_aug_s_norm}
\frac{\partial \mathcal{L}_{\text{aug}}}{\partial s_i} \approx  \frac{\exp{(\lambda (d_i - d_{\text{avg}}) + s_i)}}{\sum_i \exp{(\lambda (d_i - d_{\text{avg}}) + s_i)}}.
\vspace{-1mm}
\end{equation}
without changing its numerical value (for small $\lambda$).
There is no difference between (\ref{eq:d_aug_s_round}) and (\ref{eq:d_aug_s_norm}) for a small $\lambda$; in any case the change is equivalent to multiplying the loss function 
by a constant that is very close to 1. 
As softmax is invariant under translation,
% When written in this form we can see that the term 
$\mathcal{L}_{\text{delay}}$ actually makes no difference to the expression as it cancels,
so (\ref{eq:d_aug_s_norm}) can be written as:
\begin{equation}
\vspace{-1mm}
\label{eq:d_aug_s_final}
\frac{\partial \mathcal{L}_{\text{aug}}}{\partial s_i} \approx \frac{\exp{(\lambda d_i + s_i)}}{\sum_i \exp{(\lambda d_i + s_i)}}.
\vspace{-1mm}
\end{equation}
Therefore, we can get these path derivatives of the augmented objective function $\mathcal{L}_{\text{aug}}$, by simply computing the regular transducer loss \eqref{eq:rnnt_L} with the modified inputs:
\begin{equation}
\vspace{-1mm}
\label{eq:s_modified}
s'_i = \lambda d_i + s_i.
\vspace{-1mm}
\end{equation}

Let $\mathbf{\pi}=\{\pi_u\}_0^{U-1}$ be the frame indexes that emit tokens $\mathrm{y}_{0...U-1}$. As we want the alignments with a lower delay to have a larger delay score, we define $d_i$ as the sum of offsets relative to the middle frame in each utterance: 
\begin{equation} 
\vspace{-1mm}
 d_i = \sum_u \left(\frac{T-1}{2} - \pi_u\right).
 \vspace{-1mm}
\end{equation}
Adding the middle-frame offset will make no difference to the derivatives; it is done to prevent the delay-penalty from changing
the numerical value of the loss function too much, which would make diagnostics harder to interpret.
As shown in Figure~\ref{fig:lattice}, we can equivalently implement \eqref{eq:s_modified} by adding the offsets on the log-probabilities of emitting non-blank tokens $y(t, u-1)$ according to the specific frame indexes $0 \le t < T$: 
\begin{equation} 
 \vspace{-1mm}
\label{eq:y_add_lambda}
 y'(t, u) = y(t, u) + \lambda \times \left(\frac{T-1}{2} - t\right).
  \vspace{-1mm}
\end{equation}

Therefore, by replacing $y(t, u)$ with $y'(t, u)$ in \eqref{eq:rnnt_forward}, it would encourage low-delay alignments while maximizing the total log-probability $\mathcal{L}$, to prevent the transducer from avidly enhancing the high-delay alignments to access more future contexts~\footnote{An alternative way to implement (\ref{eq:s_modified}) is to apply the delay penalty on log-probability of emitting blank tokens $\varnothing(t, u)$ in opposite direction.}. As shown in the red line in Figure~\ref{fig:training_delay}, by applying the delay penalty on transducer, we can gradually achieve a lower symbol delay for the streaming Conformer.

\label{sec:method}

% \begin{comment}
% \section{algorithm}
% \input{algorithm}
% \label{sec:algorithm}
% \end{comment}

% \begin{comment}
% \section{Latency Metrics}
% \input{metrics}
% \label{sec:metrics}
% \end{comment}

\presec
\section{Experiments}
\postsec

\subsection{Latency metrics}
We measure the latency of streaming models with two types of delay metrics described below: (1) Mean Alignment Delay (MAD) and (2) Mean End Delay (MED). The ground-truth word-time alignments are obtained by performing forced alignment with the Montreal Forced Aligner tool \footnote{https://github.com/MontrealCorpusTools/Montreal-Forced-Aligner}. For simplicity, we only consider the correctly recognized words for both metrics. Specifically, MAD is the mean of word time difference between the predicted alignments and ground truth, which is defined as:
\begin{equation}
\vspace{-1mm}
    \text{MAD} \coloneqq \frac{1}{\sum_{n=0}^{N-1} S_n} \sum_{n=0}^{N-1} \sum_{s=0}^{S_n - 1} \left(\hat{t}_s^n - t_s^n\right)
\vspace{-1mm}
\end{equation}
Herein, $\hat{t}_s^n$ and $t_s^n$ are the timestamps of the $s$-th word in prediction and ground truth respectively. $N$ is the number of utterances. $S_n$ is the number of matched words between prediction and reference in the $n$-th utterance. MED only considers the emitting time of the last word in an utterance, which is calculated as:
\begin{equation}
\vspace{-1mm}
    \text{MED} \coloneqq \frac{1}{N}\sum_{n=0}^{N-1}\left(\hat{t}_{end}^n - t_{end}^n\right),
\vspace{-1mm}
\end{equation}
where $\hat{t}_{end}^n$ and $t_{end}^n$ are the timestamps of the last word in prediction and ground truth respectively.
\presec
\subsection{Experimental Setup}
\postsec
Our experiments are conducted on the popularly used open-source dataset LibriSpeech~\cite{librispeech}, containing 1000 hours of English reading speech.
%We train the models on \textit{train-clean-100}, \textit{train-clean-360}, and \textit{train-other-500}, and perform evaluation on \textit{test-clean} and \textit{test-other}.
We employ Lhotse~\cite{Zelasko_Lhotse_a_speech_2021} for data preparation. The acoustic features are 80-dimension Mel filterbank with a frame length of 25 ms and frame shift of 10 ms. SpecAugment~\cite{SpecAugment} and noise augmentation based on MUSAN~\cite{musan2015} are applied during training to improve generalization capability. 
% We also use noise augmentation by randomly mixing the noise audios from MUSAN~\cite{musan2015} into the training data with a SNR ranging from 10 to 20. 
Furthermore, speed perturbation~\cite{speed_perturb} with factors 0.9 and 1.1 are used to triple the training set. The transcripts are tokenized into 500-class word pieces with Byte Pair Encoding (BPE)~\cite{bpe}. 

To evaluate the effectiveness and robustness of the proposed method, we adopt streaming Conformer~\cite{anmol2020conformer} and  unidirectional LSTM as encoders respectively. Both of the Conformer and LSTM consist of 12 layers, where a convolutional downsampling layer with a factor of 4 is first used to obtain the 512-dimension feature embedding. Similar to~\cite{zhang2020unified}, we train the Conformer with block-triangular masks to limit the future context within a dynamic chunk size and infer it with a fixed chunk size of 640 ms. For each of Conformer encoder layer, the attention dimension and the feed-forward dimension are 512 and 2048, respectively. The LSTM layers adopt similar residual connection structure as in Conformer~\cite{anmol2020conformer}, each of which is composed of a unidirectional LSTM layer with 1024 hidden units and a feed-forward layer with a hidden dimension of 2048. We use a stateless decoder~\cite{ghodsi2020rnntstateless}, which consists of an embedding layer followed by a 1-D convolutional layer with a kernel size of 2. Pruned transducer loss~\cite{pruned-rnnt} is adopted for low memory usage and efficient computation. The delay penalty is applied on both the simple loss and pruned loss. We conduct experiments with 5 values of $\lambda$ in \eqref{eq:y_add_lambda}, including 0.0015, 0.0030, 0.0060, 0.0075, 0.0100. 
% The mechanisms in delay penalty and FastEmit~\cite{fastemit} are applied on both of the simple loss and pruned loss. 

\begin{figure}[ht]
\centering
\subfigure[MAD-WER trade-off]{
\centering
    \begin{tikzpicture}[scale=0.6]
        \begin{axis}[
            xlabel style = {font=\large},
            xlabel={WER (\%)},
            ylabel style = {font=\large, yshift=0pt},
            ylabel={MAD (ms)},
            xmin=5.9, xmax=8.5,
            ymin=95, ymax=670,
            ticklabel style = {font=\large},
            ymajorgrids=true,
            grid style=dashed,
            legend pos=north east,
            legend cell align={left},
            % legend style={
            %   font=\large
            % },
        ]
        \addplot+ [
            every mark/.append style={scale=1.2}]
            coordinates {
            (6.03,614)(6.21,466)(6.53,360)(6.65,258)(7.13,195)(7.56,162)
            };
        \addplot+ [
            every mark/.append style={scale=1.2}]
            coordinates {
            (6.45,470)(6.7,341)(6.99,254)(7.15,185)(7.57,143)(7.89,119)
            };
        \addplot+ [
            every mark/.append style={scale=1.2}]
            coordinates {
            (6.94,408)(7.18,292)(7.54,219)(7.69,169)(7.98,139)(8.34,117)
            };
        \legend{chunk=640, chunk=320, chunk=160}
        \end{axis}
    \end{tikzpicture}}
\subfigure[MED-WER trade-off]{
\centering
    \begin{tikzpicture}[scale=0.6]
        \begin{axis}[
            xlabel style = {font=\large},
            xlabel={WER (\%)},
            ylabel style = {font=\large, yshift=0pt},
            ylabel={MED (ms)},
            xmin=5.9, xmax=8.5,
            ymin=130, ymax=765,
            ticklabel style = {font=\large},
            ymajorgrids=true,
            grid style=dashed,
            legend pos=north east,
            legend cell align={left},
            % legend style={
            %   font=\large
            % },
        ]
        \addplot+ [
            every mark/.append style={scale=1.2}]
            coordinates {
            (6.03,725)(6.21,547)(6.53,432)(6.65,323)(7.13,258)(7.56,221)
            };
        \addplot+ [
            every mark/.append style={scale=1.2}]
            coordinates {
            (6.45,557)(6.7,399)(6.99,307)(7.15,235)(7.57,191)(7.89,167)
            };
        \addplot+ [
            every mark/.append style={scale=1.2}]
            coordinates {
            (6.94,468)(7.18,338)(7.54,267)(7.69,215)(7.98,181)(8.34,159)
            };
        \legend{chunk=640, chunk=320, chunk=160}
        \end{axis}
    \end{tikzpicture}}
\caption{\small{Delay-accuracy trade-off comparison using different decoding chunk sizes (ms) for streaming Conformer.}}
\label{fig:trade_off_chunk}
\vspace{-4mm}
\end{figure}
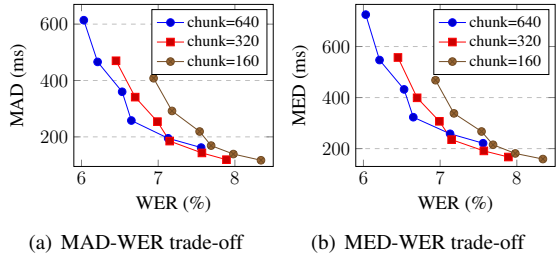

\begin{figure}[ht]
\centering
\subfigure[MAD-WER trade-off]{
\centering
    \begin{tikzpicture}[scale=0.6]
        \begin{axis}[
            xlabel style = {font=\large},
            xlabel={WER (\%)},
            ylabel style = {font=\large, yshift=0pt},
            ylabel={MAD (ms)},
            xmin=6.1, xmax=7.65, ymin=-100, ymax=246,
            ticklabel style = {font=\large},
            ymajorgrids=true,
            grid style=dashed,
            legend pos=north east,
            legend cell align={left},
        ]
        \addplot+ [
            every mark/.append style={scale=1.2}]
            coordinates {
            (6.29,203)(6.44,103)(6.81,17)(7.28,-28)(7.42,-42)
            };
        \addplot+ [
            every mark/.append style={scale=1.2}]
            coordinates {
            (6.21,226)(6.525,120)(6.65,18)(7.13,-45)(7.56,-78)
            };
        \legend{fastemit,delay-penalty}
        \end{axis}
    \end{tikzpicture}}
\subfigure[MED-WER trade-off]{
\centering
    \begin{tikzpicture}[scale=0.6]
        \begin{axis}[
            xlabel style = {font=\large},
            xlabel={WER (\%)},
            ylabel style = {font=\large, yshift=0pt},
            ylabel={MED (ms)},
            xmin=6.1, xmax=7.65, ymin=-40, ymax=327,
            ticklabel style = {font=\large},
            ymajorgrids=true,
            grid style=dashed,
            legend pos=north east,
            legend cell align={left},
        ]
        \addplot+ [
            every mark/.append style={scale=1.2}]
            coordinates {
            (6.29,282)(6.44,177)(6.81,92)(7.28,45)(7.42,27)
            };
        \addplot+ [
            every mark/.append style={scale=1.2}]
            coordinates {
            (6.21,307)(6.525,192)(6.65,83)(7.13,18)(7.56,-19)
            };
        \legend{fastemit,delay-penalty}
        \end{axis}
    \end{tikzpicture}}
\caption{\small{Delay-accuracy trade-off comparison on Conformer.}}
\label{fig:trade-off-comformer}
\vspace{-4mm}
\end{figure}
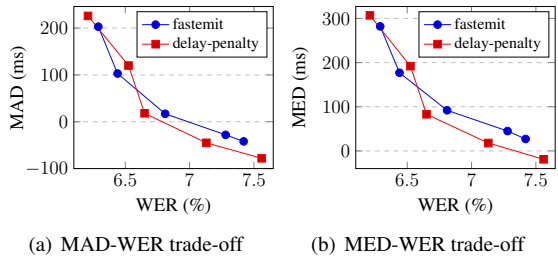

\begin{figure}[ht]
\centering
\subfigure[MAD-WER trade-off]{
\centering
    \begin{tikzpicture}[scale=0.6]
        \begin{axis}[
            xlabel style = {font=\large},
            xlabel={WER (\%)},
            ylabel style = {font=\large, yshift=0pt},
            ylabel={MAD (ms)},
            xmin=6.72, xmax=8.15,
            ymin=150, ymax=337,
            ticklabel style = {font=\large},
            ymajorgrids=true,
            grid style=dashed,
            legend pos=north east,
            legend cell align={left},
            % legend style={
            %   font=\large
            % },
        ]
        \addplot+ [
            every mark/.append style={scale=1.2}]
            coordinates {
            (6.89,309)(7.12,264)(7.40,212)(7.81, 180)(8.05,160)
            };
        \addplot+ [
            every mark/.append style={scale=1.2}]
            coordinates {
            (6.82,327)(6.97,271)(7.32,222)(7.72,188)(7.97,163)
            };
        \legend{fastemit, delay-penalty}
        \end{axis}
    \end{tikzpicture}}
\subfigure[MED-WER trade-off]{
\centering
    \begin{tikzpicture}[scale=0.6]
        \begin{axis}[
            xlabel style = {font=\large},
            xlabel={WER (\%)},
            ylabel style = {font=\large, yshift=0pt},
            ylabel={MED (ms)},
            xmin=6.72, xmax=8.15,
            ymin=192, ymax=370,
            ticklabel style = {font=\large},
            ymajorgrids=true,
            grid style=dashed,
            legend pos=north east,
            legend cell align={left},
            % legend style={
            %   font=\large
            % },
        ]
        \addplot+ [
            every mark/.append style={scale=1.2}]
            coordinates {
            (6.89,348)(7.12,298)(7.40,258)(7.81, 224)(8.05,202)
            };
        \addplot+ [
            every mark/.append style={scale=1.2}]
            coordinates {
            (6.82,360)(6.97,308)(7.32,262)(7.72,227)(7.97,202)
            };
        \legend{fastemit, delay-penalty}
        \end{axis}
    \end{tikzpicture}}
\caption{\small{Delay-accuracy trade-off comparison on LSTM.}}
\label{fig:trade-off-lstm}
\vspace{-3mm}
\end{figure}
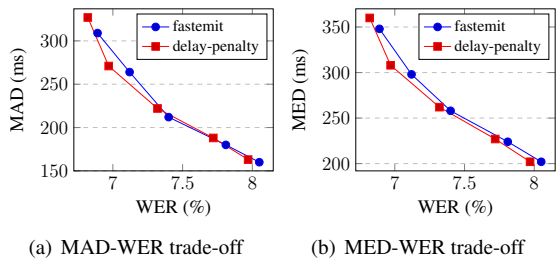

\presec
\subsection{Delay and accuracy trade-off}
\postsec

% To investigate the effect of the hyper-parameter $\lambda$ in the proposed delay penalty method on the streaming transducer model, we conduct experiments with 5 different values. 
Table~\ref{tab:librispeech} presents experimental results using different $\lambda$ in \eqref{eq:y_add_lambda}. For both Conformer and LSTM models, a larger $\lambda$ consistently leads to a lower symbol delay as well as a higher WER. It manifests that we can balance the trade-off between symbol delay and accuracy for both streaming Conformer and LSTM models in a simple and effective way by tuning $\lambda$. 
Note that for the Conformer that can access a chunk of future context, the MAD and MED is further reduced to below zero, which indicates that the model is regularized to emit symbols before they are spoken. 

We also investigate the effect of the decoding chunk size of the streaming Conformer model.  Figure~\ref{fig:trade_off_chunk} shows the delay-accuracy trade-offs with decoding chunk size of 640 ms, 320 ms, and 160 ms, respectively, where the presented results for each experiment are averaged over \textit{test-clean} and \textit{test-other}. Note that the MAD and MED here are the total delays including the latency introduced by the chunk-wise decoding, which equals half of the chunk length (i.e., 320 ms, 160 ms, and 80 ms).  The results manifest that it is preferable to decode with a larger chunk size while employing the delay penalty, which yields a better trade-off between symbol delay and accuracy. 

\presec
\subsection{Comparison with FastEmit}
\postsec

We also conduct experiments to compare our proposed delay penalization method with FastEmit~\cite{fastemit}. For FastEmit, $\lambda$ is set to 0.0030, 0.0060, 0.0100, 0.0150, and 0.0200, respectively. The FastEmit mechanism is also applied on both of the simple loss and pruned loss in pruned transducer~\cite{pruned-rnnt}. Figure~\ref{fig:trade-off-comformer} and Figure~\ref{fig:trade-off-lstm} present the delay-accuracy trade-offs of applying FastEmit~\cite{fastemit} and delay penalty as latency regularization on both Conformer model and LSTM model, respectively. For an overall comparison, the presented results for each experiment are averaged over \textit{test-clean} and \textit{test-other}. It shows that our method achieves similar delay-accuracy trade-offs to FastEmit~\cite{fastemit}, while our method provides a more detailed demonstration explaining why it is able to cause alignment time to change. 

\begin{table}[t]
\small
  \centering
  \caption{\small{ASR results on LibriSpeech using Conformer and LSTM as streaming encoder respectively.}}
  \label{tab:librispeech}
  \setlength{\tabcolsep}{4pt} % Default value: 6pt
  \renewcommand{\arraystretch}{1.2}
  \resizebox{1\linewidth}{!}{
  \begin{tabular}
  {l|c|ccc|ccc}
  \toprule
  \multirow{3}{*}{Method} & \multirow{3}{*}{$\lambda$} & \multicolumn{3}{c|}{test-clean} & \multicolumn{3}{c}{test-other} \\
  & & WER & MAD & MED & WER & MAD & MED \\
  & & (\%) & (ms) & (ms) & (\%) & (ms) & (ms) \\
  \midrule
  \multirow{6}{*}{Conformer} & 0 & 3.4 & 373 & 485 & 8.66 & 374 & 484 \\
  & 0.0015 & 3.42 & 213 & 295 & 9.01 & 238 & 319 \\
  & 0.0030 & 3.70 & 102 & 176 & 9.35 & 137 &  208 \\
  & 0.0060 & 3.74 & -3 & 62 & 9.56 & 39 & 104 \\
  & 0.0075 & 4.13 & -62 & -1 & 10.13 & -27 & 37 \\
  & 0.0100 & 4.67 & -93 & -37 & 10.44 & -62 & -1 \\
  \midrule
  \multirow{6}{*}{LSTM} & 0 & 3.78 & 418 & 437 & 9.55 & 419 & 425 \\
  & 0.0015 & 3.82 & 316 & 353 & 9.82 & 337 & 366 \\
  & 0.0030 & 3.86 & 257 & 299 & 10.08 & 284 & 317 \\
  & 0.0060 & 4.11 & 206 & 250 & 10.53 & 237 & 273 \\
  & 0.0075 & 4.52 & 172  & 214 & 10.91 & 203 & 240 \\
  & 0.0100 & 4.53 & 148 & 189 & 11.40 & 178 & 214 \\
  \bottomrule
  \end{tabular} }
  \vspace{-3mm}
\end{table}

\label{sec:experiments}

\presec
\section{Conclusion}
\postsec

We propose a method of delay penalty on transducer, which is able to penalize the symbol delay in a simple and efficient way without any extra token-time alignments. We provide a detailed proof explaining why our method is able to cause alignment time to change, so as to reduce the symbol delay. We verify the proposed method on both streaming Conformer and LSTM models.
The experimental results show that we can get a promising trade-off  between symbol delay and accuracy by tuning the hyper parameter $\lambda$. 
%The experimental results show that we can easily balance the trade-off between symbol delay and accuracy by tuning the hyper parameter $\lambda$. 

%It also  experiments comparing our method with the FastEmit   also show that we can get similar sometimes better WER versus delay trade-off.
\label{sec:conclusion}

\newpage
\bibliographystyle{IEEEbib}
\bibliography{strings,refs}

\end{document}